\documentclass[12pt]{article}

\usepackage[active]{srcltx}
\usepackage{latexsym}
\usepackage{amsmath}
\usepackage{amsfonts}
\usepackage{mathrsfs}
\usepackage{dsfont}
\usepackage{amscd}
\usepackage{verbatim}
\usepackage{xcolor}
 \csname
@addtoreset\endcsname{equation}{section}

\usepackage{lipsum}
\usepackage{xargs}
\usepackage[normalem]{ulem}

\usepackage[colorinlistoftodos,prependcaption,textsize=tiny]{todonotes}
\newcommandx{\unsure}[2][1=]{\todo[linecolor=red,backgroundcolor=red!25,bordercolor=red,#1]{#2}}
\newcommandx{\change}[2][1=]{\todo[linecolor=blue,backgroundcolor=blue!25,bordercolor=blue,#1]{#2}}
\newcommandx{\info}[2][1=]{\todo[linecolor=red,backgroundcolor=red!25,bordercolor=red,#1]{#2}}
\newcommandx{\Sinfo}[1]{\todo[backgroundcolor=red!25,bordercolor=red,noline]{S:\,#1}}
\newcommandx{\Winfo}[1]{\todo[backgroundcolor=blue!25,bordercolor=blue,noline]{W:#1}}
\newcommandx{\improvement}[2][1=]{\todo[linecolor=Plum,backgroundcolor=Plum!25,bordercolor=Plum,#1]{#2}}
\newcommandx{\thiswillnotshow}[2][1=]{\todo[disable,#1]{#2}}

\def\p{\partial}

\def\a{\alpha}
\def\b{\beta}

\def\s{\sigma}

\def\ps{\psi}

\def\be{\begin{equation}}
\def\ee{\end{equation}}
\def\bea{\begin{eqnarray}}
\def\eea{\end{eqnarray}}
\def\ba{\begin{array}}
\def\ea{\end{array}}

\def\tr{\text{tr}}

\def\12{\frac{1}{2}}

\def\eNm1{\overset{\scriptscriptstyle{(N-1)}}{e}}

\begin{document}

\begin{flushright}
\end{flushright}

\vspace{25pt}

\begin{center}


{\LARGE \bf Moduli Anomalies and Local Terms in the\\
\vskip.5cm
Operator Product Expansion}

\vspace{25pt}
{\large Adam Schwimmer$^{a}$ and Stefan~Theisen$^{b}$}

\vspace{10pt}
{\sl\small
$^a${\it  Weizmann Institute of Science, Rehovot 76100, Israel}
}

\vspace{10pt}
{\sl\small
$^b${\it Max-Planck-insitut f\"ur Gravitationsphysik, Albert-Einstein-Institut, 14467 Golm, Germany}

}

\vspace{70pt} {\sc\large Abstract}\end{center}
Local terms in the Operator Product Expansion in Superconformal Theories with extended supersymmetry are identified.
Assuming a factorized structure for these terms their contributions are discussed.

\vspace{60pt}
\begin{center}
\today
\end{center}

\newpage

\tableofcontents

\section{Introduction}

Conformal Field Theories (CFT) are specified by the list of primary operators with their dimensions and the structure
constants, i.e. the normalizations of the correlators of three primary operators. Generically from this data any correlator in
the theory can be reconstructed.

In the present paper we will discuss additional input data which is needed for certain classes of primaries.
This additional data can in principle be recovered from the symmetries of the CFT. The possible need for additional data is
suggested by the presence of  ``semilocal terms" in certain
correlators. These terms were discussed recently from various points of view \cite{Bzowski,Prilepina,Nakayama}.
We will call ``semilocal" in a generic sense those terms which do not have
the full analytic structure in momentum space expected in a
correlator with  singularities in all kinematical invariants. The semilocal terms have a singularity
(logarithmic or powerlike) in one kinematical variable while the dependence on the other invariants is polynomial.
We will distinguish these terms from the ordinary local terms which are  polynomial in all invariants
and which appear in specific correlators.
In position space locality means that all operator insertion points coincide while in semi-local correlators
only some of them do. The generic situation which shows the full analytic structure in momentum space is when
they are all different.

We will use alternatively the
operatorial or the covariant (i.e. based on the analytic structure of the correlators) language 
and will verify their
compatibility in each case. We start by defining the local terms in the Operator Product Expansion (OPE) and necessary
conditions for their appearance.

In a CFT the  information about the structure constants can be reconstructed from the operator product expansion
which we will use in the following:
given two primaries ${\cal O}_1,{\cal O}_2$ with scaling dimensions $d_1$ and $d_2$, their OPE is
\begin{equation}
{\cal O}_1(x)\,{\cal O}_2(y)=c_{123} \big((x-y)^2\big)^{{1\over2}(d_3-d_1-d_2)}\,{\cal O}_3(y)\,+\,\dots
\end{equation}
Generically this is singular for $(x-y)^2\sim0$. In momentum space the singularity produces
an analytic function of the kinematical invariant  $p^2$ with a branch point at $p^2=0$,
which in general has both an imaginary and a real part. A special situation arises  when the singular function  has the form
$\partial^n \delta^{d}(x-y)$.
Then, in momentum space, we have a purely real (polynomial) dependence on $p^2$, and in configuration space
the singular function is ``local''. For such a situation to occur, the following conditions have to be satisfied:

\begin{itemize}

\item[{a)}] the dimensions should fullfil $d_1+d_2-d_3=\hbox{integer} \geq d$ (the space-time dimension),
following from the above form of the local singular function;

\item[{b)}] the ordinary structure constant/OPE coefficient accompanying the singular function
$\big((x-y)^2\big)^{-{1\over2}(d+2n)}$ which, in even dimensions,  
corresponds in momentum space to $(p^2)^{n} \log{p^2\over \Lambda^2}$,
must vanish. If this were not the case its contribution to the three point function would be the most singular
and would ``mask" the semilocal structure which contains just the polynomial.

\end{itemize}
While the momentum dependence involving ${\cal O}_1,{\cal O}_2$ will be polynomial,
the singular behaviour in the correlators is obtained by singularities of ${\cal O}_3$ with other operators.

We remark a basic difference between the ordinary terms in the OPE and the local ones.
For  the ordinary  OPE terms coming from a unique three point function there is a relation between the three possible
orderings. In contrast to this,
the local terms in the OPE for the three possible orderings are independent: they produce the three operator correlators
through convolutions with different two point functions.

A basic issue which will be important in our discussion is whether the local term in the OPE can be removed and
its effect  in the various correlators absorbed in a redefinition in the   ordinary framework of the CFT.
Alternatively this reduces to the question
whether the local term can be formulated in a universal  regularization independent way.

To illustrate this setup and the issue of universality we review the well known case of the Zamolodchikov
metric on the conformal manifold  in $d=2$. This was  studied a long time ago in \cite{Kutasov}.
We will denote moduli, which are exactly marginal operators, by  $M_i$. They  have dimension $d$, which
we assume to be even, and their structure constants
vanish by their  requirement of being moduli.

In the present paper global properties of the moduli space do not play an essential role and we will not discuss them.
We will limit the range of moduli couplings
to the vicinity of a fiducial CFT where the moduli space does not have singularities.
In this range, where the conformal deformation of the CFT by adding the moduli multiplied by couplings is well defined,
we will treat the couplings as particular $x-$independent values of the sources.
They are local coordinates
on the moduli space which is also referred to as the conformal manifold.
If the space-time dependent sources are denoted
by $J^i(x)$, the conformal deformation is obtained by adding
\begin{equation}
\delta S={1\over\pi^{d/2}}\int d^d x \sqrt{g}\,J^i(x)\,M_i(x)
\end{equation}
to the  action of the fiducial CFT.

The conditions  a) and b) are satisfied for the moduli and  we can search for the local contributions
in their OPE. In our approach these are a consequence of the type B trace anomaly in the CFT.
To be specific, consider the situation in $d=2$ \cite{Gomis}: if the  $x-$dependent sources of the moduli are
denoted by  $J^i$, the anomaly is
\begin{equation}\label{eq1}
\mathcal{A}=-{1\over 4\pi}\int d^2x \sqrt{g}\,\sigma\, G_{ij}(J)\partial_{\mu}J^i \partial^{\mu}J^j
\end{equation}
$G_{ij} (J)$ is the Zamolodchikov metric. The universal information contained in \eqref{eq1}, like in every
type B anomaly \cite{DDI,DS}, is the
logarithmically divergent counterterm \begin{equation}\label{eq2}
{1\over 8\pi}\log(\Lambda^2)\int d^2x\sqrt{g}\, G_{ij}(J)\partial_{\mu}J^i \partial^{\mu}J^j
\end{equation}
which contains  semilocal correlators of the moduli, the singularity in momentum space being 
inherited from the two point function. 

Taking three functional derivatives  with respect to $J$ and Fourier transforming,  we obtain
\begin{equation}\label{eq3}
\langle M_i(p_1)M_j(p_2)M_k(p_3)\rangle
={\pi^2\over 4}\log(\Lambda^2)\left(p_3^2\, \Gamma_{ij,k} +\hbox{cyclic permutations}\right)
\end{equation}
Here $\Gamma$ is the Christoffel connection for the Zamolodchikov metric,
evaluated at the point in moduli space we are studying.
Since the two point function at that same point in moduli space is
\begin{equation}\label{eq4}
\langle M_i(p)M_j(-p)\rangle ={\pi\over 4} G_{ij}\, p^2 \log{p^2/\Lambda^2}
\end{equation}
the logarithmically divergent contribution in the three point function \eqref{eq3}
will be reproduced by a local term in the OPE:
\begin{equation}\label{eq5}
M_i(p_1) M_j(p_2)={\pi\over 4}\Gamma_{ij}^k M_k(p_1+p_2)
\end{equation}
as proposed in \cite{Kutasov}.
In configuration space this corresponds to
\begin{equation}\label{eq6}
M_i(x)M_j(y)={\pi\over4}{\delta}^2 (x-y)\Gamma_{ij}^k M_k(x)
\end{equation}

In the previous argument we used universal features of the logarithmically divergent counterterm. This will be part of our
approach, i.e. we will always start with the semilocal term in the  correlator which has all the symmetries and analyticity
properties of the theory
and derive from it the local contribution to the OPE needed to reproduce it. In particular the
transformation properties under  source reparametrizations reflect the covariance of the Zamolodchikov metric defined by the
counterterm.

We now discuss the  issue of the universality of the local term in the OPE above.
The key is the behaviour of this term under source reparametrization,  which is a symmetry of the theory.
The term found is normalized by the Christoffel connection of the Zamolodchikov  metric.
It transforms inhomogenously under
reparametrizations of the sources which
suggests that it is not universal. Indeed, using Gaussian normal coordinates at a given point in
moduli space, the connection can be put to zero. This shows that the semilocal term which it represents
can be obtained  from the ordinary
set up of the theory without the need  of local terms in the OPE. Moreover in \cite{Friedan} an explicit procedure in a special
regularization is given which shows how one could recover the reparametrization invariant  information contained in the
Zamolodchikov metric from correlators calculated without using additional local terms in the OPE.
An example where some local
contributions to the OPE can be removed while others not was discussed recently in \cite{Wang}.

The situation  in $d=4$ is similar:
we start from the anomaly \cite{Gomis}
\begin{equation}\label{eq8}
{\cal A}={1\over 192\pi^2}\int d^4 x\sqrt{g}\,\s\left( G_{ij}(J)\,\hat\square J^i\,\hat\square J^j
-2 G_{ij}(J)\p_\mu J^i\left( R^{\mu\nu}-{1\over3}g^{\mu\nu}R\right)\p_\nu J^j\right)
\end{equation}
where $\hat\square J^i=\square J^i+\Gamma^i_{jk}\nabla^\mu J^j\,\nabla_\mu J^j$.
This reflects the counterterm (in flat space-time) proportional to
\begin{equation}
\log(\Lambda^2) \int d^4 x\,G_{ij}(J)\,\hat\square J^i\,\hat\square J^j
\end{equation}
and one gets again \eqref{eq3} through \eqref{eq6}, with the only difference (apart from the numerical factor)
that $p^2$ is replaced by $(p^2)^2$ and one has
a four-dimensional $\delta$-function in \eqref{eq6}. The generalization of these equations to arbitrary
even dimensions is obvious\footnote{but not so the generalization of \eqref{eq8} which can, however, be
worked out case by case.}. Again, the apparent addition to the OPE specified above is removable and therefore not universal.

In this paper we will apply the same logic to identify  new local contributions to the OPE involving currents and moduli.
However, the terms which we single out cannot be removed and are therefore universal.
This is a consequence of supersymmetry\footnote{In the present paper we use extended supersymmetry, 
but some results may also follow from  e.g. ${\cal N}=1$ superconformal symmetry  in $d=4$.
We thank Z. Komargodski for pointing out this possibility.} 
i.e. we will study ${\cal N}=(2,2)$ superconformal theories in $d=2$ and ${\cal N}=2$ SCFTs in $d=4$.
The common feature we will find is that the local terms in the OPE will
be normalized by the  Zamolodchikov metric itself and therefore cannot be removed by source reparametrization.

Once the additional terms in the OPE are identified we study whether  factorization can be used for the local terms in a
manner analogous to the decomposition in terms of the ordinary structure constants. The local terms generate classes
of contributions to certain
correlators of the theory in addition to the usual one. We check various requirements for these contributions, in particular their
consistency with  supersymmetry. While for $(2,2)$ theories in $d=2$ we have a complete and consistent construction,  in $d=4$
we have to face an ambiguity in the separation of certain correlators into
ordinary  contributions (the result of combining the usual three point structures) and the ones produced through
factorization from the local terms.

The paper is organized as follows:
in Section 2 we will review the structure of moduli anomalies for
${\cal N}=2$ superconformal theories in four dimensions and identify the logarithmically divergent counterterm involving currents.
We identify a contribution to the trace anomaly originating in the correlator of a $U(1)_{\cal R}$ current and moduli and 
applying the logic outlined
above we identify the local term in the OPE of two moduli giving the current.

In Section 3 we discuss ${\cal N}=(2,2)$ superconformal theories in $d=2$. In the general expression for the 
superconformal anomalies
we identify a contribution to the $U(1)$ anomaly  and we determine the local contribution in the OPE needed to reproduce it.
Using special features of $d=2$ we construct the full anomalous part of the effective action which incorporates terms obtained
through factorization from  the local additions to the OPE.

In Section 4, where we present our conclusions, we discuss the possibility of using the local terms in the
OPE in  a factorized manner, the consequences of such an assumption and the consistency checks needed.

In Appendix A we give a proof for the absence of ordinary structure constants of moduli and conserved currents.
In Appendix B we work out explicitly the local contributions for the simplest example of an ${\cal N}=2$ theory in $d=4$:
the free Maxwell gauge supermultiplet. We identify in a Feynman diagram calculation the contribution of redundant operators
 which leads to the semilocal structures we find. 
This suggests that in a diagrammatic calculation the local terms in the OPE 
can be replaced by redundant (i.e. vanishing on shell) operators.  We check the consistency of these factorized contributions
by calculating the  anomalous four moduli correlator consistent with supersymmetry  in this model as a sum of factorized 
and ordinary contributions.
In Appendix C we work out a similar field theoretic model realizing the structures we found for the 
${\cal N}=(2,2)$ theories in two dimensions.

\section{${\cal N}=2$ in $d=4$}

In this section we discuss  four-dimensional ${\cal N}=2$ superconformal theories with moduli. Superconformal theories in
$d=2$ with $(2,2)$ supersymmetry, which have several special, simplifying features, will be discussed in the next section.

Conformal field theories with ${\cal N}=2$ supersymmetry have an
$SU(2)\times U(1)$ ${\cal R}$-symmetry, of which the $U(1)$ factor is anomalous. 
The basic result of this section is that this anomalous $U(1)$ ${\cal R}$-current appears as a local term in the
$M\,\overline M$ operator product. This follows from the structure  of
a counterterm related to a type B Weyl-anomaly.
This counterterm is required by supersymmetry.
In Appendix B we will verify some of our general results and claims by
looking at pure  ${\cal N}=2$ $U(1)$ gauge theory where all features appear at one loop order and can be explicitly
computed.

We  gauge the global symmetries
and couple the CFT to an external metric and gauge fields, the sources for the energy momentum tensor 
and for the ${\cal R}$-symmetry currents, respectively.  
And, of course, we also have the sources for the moduli.
The anomalous Ward identities are then most succinctly incorporated in the effective action,
which is the non-local functional of the sources obtained by integrating out the CFT. 
It necessarily violates some of the symmetries and the anomaly ${\cal A}$ is 
the variation of the generating functional under these transformations. Here the anomalous symmetries are 
super-Weyl transformations, which are parametrized by a chiral superfield $\Sigma$, whose lowest component 
$\Sigma|=\sigma+i\,\alpha$ parametrizes Weyl ($\sigma$) and $U(1)_{\cal R}$ transformations ($\alpha$). 

The super-Weyl anomaly is therefore the variation of the effective action with the chiral superfield parameter 
$\Sigma$ \cite{Kuzenko,Gomis}
\begin{equation}
\begin{aligned}\label{anomaly-superspace}
{\cal A}&={1\over 16\pi^2}\Big(\int d^4 x\,d^4\theta\,{\cal E}\,\Sigma \big(a\,\Xi+(c-a)W^{\a\b}W_{\a\b}\big)+\hbox{c.c.}\Big)\\
&\qquad\qquad\qquad+{1\over 192\pi^2}\int d^4 x\,d^4\theta\,d^4\bar\theta\,E\,(\Sigma+\bar\Sigma) K(J,\bar J)
\end{aligned}
\end{equation}
$c$ and $a$ are the Weyl anomaly coefficients which are characteristic of a particular SCFT. The normalization
of the last term is fixed by the two-point function of the moduli, which is given by the Zamolodchikov metric for which
$K$ is the K\"ahler potential.
The component expansion of the above expression is \cite{deWit,Gomis}\footnote{In \cite{Gomis}
the $U(1)_{\cal R}$ and $SU(2)_{\cal R}$ gauge fields were set to zero.}
\begin{equation}\label{anpol4}
\begin{aligned}
{\cal A}&
={1\over16\pi^2}\int\!\! d^4 x\sqrt{g}\bigg\lbrace -a\,\s\left(E_4-{2\over3}\,\square R\right)
+c\,\s\,C^{\mu\nu\rho\s}C_{\mu\nu\rho\s}-2\,c\,\s\, F^{\mu\nu}F_{\mu\nu}
+{1\over2}c\,\s\,\tr\big(H^{\mu\nu}H_{\mu\nu}\big)\\
&\qquad\qquad\qquad+(a-c)\,\a\, R^{\mu\nu\rho\s}\tilde R_{\mu\nu\rho\s}
+2(c-a)\,\a\, F_{\mu\nu}\tilde F^{\mu\nu}
+{1\over 2}(2\,a-c)\,\a\,\tr\big(H_{\mu\nu}\tilde H^{\mu\nu}\big)\\
&\qquad\qquad\qquad+4\,a\,\nabla^\mu A_\mu\,\square\a-8\, a\, A^\mu\Big(R_{\mu\nu}-{1\over3}R\, g_{\mu\nu}\Big)\nabla^\nu\a
-8\,a\,F_{\mu\nu}\,A^\mu\,\nabla^\nu\s\bigg\rbrace\\
&+{1\over96\pi^2}\int\!\! d^4x\sqrt{g}\,\bigg\{
\!\sigma {\cal R}_{i\bar k j\bar l}\nabla^\mu J^i\nabla_\mu J^j\,\nabla^\nu\bar J^k\nabla_\nu \bar J^l
\!+\!\sigma\,G_{i\bar\jmath}\!\left(\!\hat\square J^i\,\hat\square\bar J^j
\!-\!2\left(R^{\mu\nu}\!-\!{1\over3}R\,g^{\mu\nu}\!\right)\p_\mu J^i\,\p_\nu\bar J^j\right)\\
&+{1\over2}K\,\square^2\s+{1\over6}K\,\p^\mu R\,\p_\mu\s+K\left(R^{\mu\nu}-{1\over3}R\,g^{\mu\nu}\right)\nabla_\mu\nabla_\nu\s
-2\,G_{i\bar\jmath}\,\nabla^\mu J^i\,\nabla^\nu \bar J^j\,\nabla_\mu\nabla_\nu\s\\
&+i\,G_{i\bar\jmath}\left(\hat\nabla^\mu\hat\nabla^\nu J^i\,\nabla_\nu\bar J^j
-\hat\nabla^\mu\hat\nabla^\nu\bar J^j\,\nabla_\nu J^i\right)\p_\mu\a
-\nabla^\mu{\cal A}_\mu\,\square\a+2\,{\cal A}^\mu\left(R_{\mu\nu}-{1\over3}R\,g_{\mu\nu}\right)\!\p^\nu\a\\
&- \s\,F_{\mu\nu}\,{\cal F}^{\mu\nu}
+2\,F_{\mu\nu}\,{\cal A}^\mu\,\nabla^\nu\s+F_{\mu\nu}\,\nabla^\mu K\,\nabla^\nu\a\bigg\}
\end{aligned}
\end{equation}
Here ${\cal A}_\mu$ is the K\"ahler connection, defined as
\begin{equation}\label{defA}
{\cal A}_\mu={i\over2}\Big(\p_j K\,\p_\mu J^j-\p_{\bar\jmath}K\,\p_\mu\bar J^{\bar\jmath}\Big)
\end{equation}
and ${\cal F_{\mu\nu}}$ its
field strength which depends only on the K\"ahler metric and is therefore invariant under K\"ahler
shifts and covariant under holomorphic coordinate changes on the conformal manifold.
$F$ is the field strength of the $U(1)$ gauge field $A$ and $H$ that of the $SU(2)$ gauge field.  
Were it not for supersymmetry, many terms in the component
expression would be cohomologically trivial and could be dropped,
but as it is obvious from the (three irreducible) superspace expressions, supersymmetry
demands that they accompany the cohomologically non-trivial terms. Supersymmetry is also responsible for the
appearance of the target space Riemann tensor in the fourth line. In a bosonic theory it could be replaced
by an arbitrary tensor with the correct symmetries and would still be a non-trivial solution to the
Wess-Zumino consistency condition. But ${\cal N}=2$ supersymmetry requires that this tensor is the
Riemann tensor. In the general form this anomaly first appeared in \cite{OsbornAnomaly} and we
therefore refer to it as the Osborn anomaly.

One can write down a non-local action, both in superspace
and in components, whose super-Weyl variation
reproduces \eqref{anomaly-superspace}, but one is faced with the same problem as for the ordinary
Weyl anomaly in four dimensions that this ``Riegert" action does not have the correct analyticity properties 
\cite{Erdmenger}. It therefore
differs from the unknown `true' effective action by unknown non-local Weyl invariant terms.

The anomaly polynomial is invariant under a combined (field dependent)
super-Weyl transformation and a K\"ahler
transformation if their parameters are related as \cite{Gomis}
\begin{equation}\label{gauge-Kaehler}
\tilde\Sigma={1\over 24 a}F
\end{equation}
This is easy to verify for the superspace action \eqref{anomaly-superspace} and can also be verified for the
component expression \eqref{anpol4}. It is readily observed that every term in the first three lines with a bare
gauge field $A_\mu$, i.e. not appearing in the gauge invariant combination $F_{\mu\nu}$,
has a counterpart in the last two lines if we replace $A_\mu\to -{1\over 24 a}{\cal A}$. This reflects the
invariance of  $A_\mu-{1\over 24\,a}{\cal A}_\mu$ under a joint gauge and K\"ahler
transformation with  \eqref{gauge-Kaehler}.
In the ``Conclusions" section we will reformulate this symmetry in terms of the K\"ahler shift variation of the effective action.

Let us now analyse the anomaly polynomial \eqref{anpol4} further. Consider the first two terms in the
last line. The second one vanishes for $\s={\rm const.}$ and therefore will not contribute to the following argument.
The first term is a type B Weyl-anomaly and
corresponds to a counterterm, in the same way as was described in the Introduction. As such it
contains information about non-local terms in correlation functions. Taking functional derivatives with respect to
$J^i$, $\bar J^{\bar\jmath}$ and $A_\mu$, one finds the correlator
\begin{equation}
\langle M_i(k_1)\,\overline M_{\bar\jmath}(k_2)\,j_\mu(-k_1-k_2)\rangle
=-{\pi^2\over 192} G_{i\bar\jmath}(q^2\,r_\mu-q\cdot r\,q_\mu)\log\Lambda^2
\end{equation}
where $G_{i\bar\jmath}$ is evaluated for constant sources.  We have defined $q=k_1+k_2$ and
$r=k_1-k_2$.  Of course the same counterterm also
generates correlation functions of one current and an arbitrary number of moduli, always via the
current-current two-point function.
The term cannot come from an ordinary three point function since, as we show in Appendix A, the moduli being neutral
under $U(1)_{\cal R}$  the structure constant vanishes.
This  indicates that the $U(1)$ ${\cal R}$-current $j_\mu$ appears in a contact term
in the $M\,\overline M$ operator product. Since it is proportional to the Zamolodchikov metric
it cannot be removed by a reparametrization of the conformal manifold. The fact that the Zamolodchikov metric
appears is a consequence of supersymmetry.
If it were not for supersymmetry, the counterterm which is responsible for this correlator
could be omitted and one could adopt a scheme where
there are no local terms in the operator product of two moduli.

If we normalize the $U(1)_{\cal R}$
current such that the coupling in the microscopic theory is normalized to $\int A_\mu\,j^\mu$,
we find the local terms in the $M\overline M$ OPE
\begin{equation}
M_i(x)\,\overline M_{\bar\jmath}(y)\sim {\pi^4\over 48 c}G_{i\bar\jmath}\Big(\p_\mu^{(x)}\delta^4(x-y)\,j^\mu(y)
-\p_\mu^{(y)}\delta^4(x-y)\,j^\mu(y)\Big)+\dots
\end{equation}
There could be  other local terms in this operator product, but they do not
contribute to the three point function with the ${\cal R}$-symmetry current. We will give their specific form for the
particular case of the free Maxwell theory
in Appendix B.

Once we formulated the local term in operatorial language we can translate it into a covariant one: the  local term in the
OPE will give a contribution to any correlator involving moduli by coupling the moduli to the $U(1)_{\cal R}$ current.
The correlators of ${\cal R}$-currents are represented by terms in the effective action containing its source $A_{\mu}$.
Therefore the contribution of the local term in the OPE to correlators with moduli is obtained by replacing
$A_{\mu}$ in any term in the generating functional by ${1\over 24c}\cal{A}_{\mu}$. This is the general formulation of
factorization we are using.
The normalization follows from comparing the last term in the last line with the
third term in the first line of \eqref{anpol4}.

One might wonder to what extend factorization determines the form of the anomaly polynomial.
An explicit calculation in ${\cal N}=2$ super-Maxwell theory shows that the counterterm in \eqref{anpol4}
which involves the Riemann tensor on the conformal manifold (the ``Osborn anomaly"), is not completely accounted
for by factorization. The
same calculation however shows that without the local term in the above operator product the
Riemann tensor would not appear, but it is required by ${\cal N}=2$ supersymmetry.

More generally if the anomaly polynomial were given completely by factorization all the terms would contain the combination
${A}_\mu+{1 \over 24c} \cal{A}_{\mu}$.
This is clearly not the case for  \eqref{anpol4}
which contains invariant field strengths of $A_\mu$ without
the corresponding terms constructed from ${\cal A}_\mu$. This seems to be dictated by supersymmetry, because
there is no way to supersymmetrize e.g. $\a\,{\cal F}_{\mu\nu}\tilde{\cal F}^{\mu\nu}$.\footnote{At least as long as we use
only chiral multiplets to represent the sources $J^i$. This might be reminiscent of the situation in $d=2$ where
the coupling of chiral multiplets to a target space $B$-field cannot be accomplished off-shell. To do this one has to
use semi-chiral multiplets. We did not pursue the generalization of this possibility to $d=4$.}  We will elaborate on
this point in the concluding section, but already draw the partial conclusion that while the factorized contributions
of the moduli are needed,
the typical situation is that they come together with ordinary contributions obtained ignoring the local terms.
An interesting connection appears in the explicit example discussed in Appendix B: the local terms in the OPE can be
replaced by including in the moduli ``redundant operators" in the covariant calculation of the correlators. Here ``redundant"
means operators in a lagrangian CFT which vanish if one uses the equations of motion.

\section{${\cal N}=(2,2)$ in $d=2$}\label{d=2}

We begin with a review of the basic features of moduli anomalies in ${\cal N}=(2,2)$ superconformal theories \cite{Gomis}.
Extended supersymmetry implies additional global symmetries  which in this case  are the
$U(1)_A\times U(1)_V$ ${\cal R}$-symmetries.
We can choose to preserve
either one of the two $U(1)$ factors.  The second factor then belongs to the multiplet of anomalous currents.
Due to this choice we have two possible types of theories.

For concreteness, we will only discuss the theory which preserves the $U(1)_A$ ${\cal R}$-symmetry. In this case
the anomaly is\footnote{For simplicity we only consider
chiral primary moduli. For the general case, which includes also twisted chiral primary moduli, we refer to
\cite{Gomis}, where further details of the notation can also be found.}
\begin{equation}\label{anomaly:d=2}
{\cal A}=-{1\over 2\pi}\int d^2 x\Big(\sigma\,G_{i\bar\jmath}\,\p_\mu J^i\,\p^\mu \bar J^{\bar\jmath}
-{1\over2}\square\s\,K+\alpha\,\p^\mu{\cal A}_\mu+{c\over12}\big(\sigma\,R+\a\,\epsilon^{\mu\nu}F_{\mu\nu}\big)\Big)
\end{equation}
In this expression, whose superspace version will be given later,
$K$ is the K\"ahler potential on the conformal manifold, a real function of the
sources $J$ and $\bar J$ and
\begin{equation}\label{defA}
{\cal A}_\mu={i\over2}\Big(\p_j K\,\p_\mu J^j-\p_{\bar\jmath}K\,\p_\mu\bar J^{\bar\jmath}\Big)
\end{equation}
is again the K\"ahler connection. $F_{\mu\nu}$ is the field strength of the $U(1)_A$ gauge field $A_\mu$. Under local
$U(1)_V$ transformations it transforms as $\delta A_\mu=\epsilon_{\mu}{}^\nu\p_\nu\a$.
If we define $V_\mu=\epsilon_\mu{}^\nu A_\nu$, $\delta V_\mu=\p_\mu\a$ and
$\epsilon^{\mu\nu}F_{\mu\nu}=2\,\nabla^\mu V_\mu$. $\s$ parametrizes
local Weyl transformations and $c$ is the Virasoro central charge\footnote{Diffeomorphism invariance requires $c=c_L=c_R$, i.e.
absence of a gravitational anomaly.}. The relative coefficients in \eqref{anomaly:d=2} are
dictated by supersymmetry. The invariance of the effective action (nonlocal and local terms) under the axial ($U(1)_A$)
gauge transformation $\delta A_\mu=\p_\mu\beta$  is part of the definition of the theory. As a consequence
the terms  $\alpha\,\p^\mu{\cal A}_\mu$ and  $\a\,\epsilon^{\mu\nu}F_{\mu\nu}$, which can be obtained by the gauge variation of
$ A_{\mu} {\cal A}^{\mu}$ and $A_{\mu}A^{\mu}$ respectively, remain cohomologically nontrivial since the addition of these
local terms to the effective action would violate the $U(1)_A$ symmetry, i.e. the definition of the theory.
Note also a very special feature of $d=2$: the chiral anomaly can be seen not only in odd parity correlators
like in all even dimensions but also as a ``clash" between conservation in the even parity vector-vector and axial-axial
correlators when the vector and axial currents are related by a duality transformation.

We now analyse the  $\alpha\,\p^\mu{\cal A}_\mu$ term of the anomaly polynomial \eqref{anomaly:d=2}.
It represents an anomaly in the
correlator of the $U(1)_V$ current and at least one modulus and one antimodulus or, equivalently, a non-invariance
of the corresponding terms in the effective action under a vector gauge transformation of the gauge field $A_{\mu}$.
The momentum space structure of
the term which leads  to this anomaly is ${q_{\mu} q_{\nu}}\over {q^2}$, where $q$ is the momentum carried by the axial
current to which $A_{\mu}$ is coupled. By an argument similar to the one in $d=4$ which is discussed in Appendix A,
such a contribution cannot come from a modulus-antimodulus-current coupling since the
moduli are neutral under the R-symmetries.
In this case the semilocal structure involves not a logarithm  like in  $d=4$ but the characteristic $1 \over q^2$ pole
multiplied  by a polynomial in the momenta of the moduli.
In order to reproduce it we need to assume the existence of a local term in the OPE
\begin{equation}\label{postulate}
M_1(x)\,\overline M_2(y)\sim {6 \pi^2\over c}G_{i\bar\jmath}\left(\p_\mu^{(x)}\delta^2(x-y)\,\tilde\jmath^\mu(y)
-\p_\mu^{(y)}\delta^2(x-y)\,\tilde\jmath^\mu(x)\right)
\end{equation}
where $\tilde\jmath_{\mu}$ is the anomalous vector current. We could use instead equivalently the
non-anomalous axial current $j_{\mu}$ related
to $\tilde\jmath_{\mu}$ by a duality transformation.
Combining then the local term in the OPE with the correlator of two vector currents:
\begin{equation}
\langle \tilde\jmath_\mu(q)\,\tilde\jmath_\nu(-q)\rangle=\epsilon_{\mu}{}^\rho\,\epsilon_\nu{}^\s
\langle j_\rho(q)\,j_\s(-q)\rangle=-{c\over6}{q_\mu\,q_\nu\over q^2}
\end{equation}
we obtain
\begin{equation}\label{OOj}
\langle M_1(k_1)\,\overline M_2(k_2)\,\tilde\jmath_\mu(-k_1-k_2)\rangle
=-{i\over 16\,\pi^2} G_{i\bar\jmath}{q\cdot r\over q^2}q_\mu
\end{equation}

As in $d=4$ this term in the OPE cannot be removed by reparametrizations of the sources and therefore it is universal.
It leads through factorization to classes of calculable  contributions to the
effective action. These  factorized contributions  can be calculated following the
rule analogous to the one used in $d=4$ i.e.
wherever  the gauge field $A_{\mu}$ is coupled to the axial current we should replace it by the combination
$A_{\mu} + {6\over c} \epsilon_{\mu\nu} \cal{A}^{\nu}$.
This combination is manifest in the anomaly polynomial \eqref{anomaly:d=2} and 	
comparing the terms proportional to $\alpha$ 
it is clear that the anomaly involving the moduli is reproduced.

The combination
which appears is invariant under a joint transformation of the K\"ahler potential which generates
$\cal{A}_{\mu}$ by $f(J)+\bar f(\bar J)$ and a vector gauge transformation of $A_{\mu}$ with parameter
$\alpha =  -{3\,i\over c}(f-\bar f)$. The consistency between the combinations selected by factorization
and the invariance of the anomaly
polynomial  is a special feature of the two dimensional theory.

Since the local terms in the OPE are factorized we treat this effective coupling on equal terms with $A_{\mu}$,
i.e. for every term in the effective action involving $A_\mu$ we can get a term involving correlators of moduli  by the
above replacement.  As discussed in Section 2 in $d=4$ the typical situation is that both factorized local OPE contributions and
ordinary ones are needed to reproduce the total supersymmetric expressions.

In $d=2$ due the special kinematical features we are able to  give a complete  description of the anomalous part of the
effective action and to check including only the factorized contributions we get an answer consistent with supersymmetry
and the K\"ahler structure.
	
We proceed now to construct the anomalous part of the effective action. We start with the first building  block involving
the Zamolodchikov anomaly, i.e. eq.\eqref{eq1}. Generically  a type-B trace anomaly is induced by a logarithmically
divergent term, but the anomaly itself appears in the Weyl  variation of a finite
correlator which involves the sources in the divergent
term and the metric which is coupled to the energy momentum tensor \cite{DDI,DS}.
For a general type B anomaly  there are no closed expressions for  the finite part to all orders in the external sources.
Even for the standard $c$-trace anomaly in $d=4$ the finite,
non local correlator, whose Weyl variation is the anomaly, is known only in the leading order i.e. a correlator of three energy
momentum tensors \cite{Erdmenger}. For the case considered here,
i.e. the finite part reproducing the Zamolodchikov anomaly in $d=2$
containing any number of moduli and external metric perturbations, the problem can be exactly solved.
Let us start with the first
correlator which contributes to  the finite part: a correlator of two moduli and
one energy momentum. In a convenient basis the correlator has the kinematical decomposition
\begin{equation}
\begin{aligned}
\langle M_i(k_1)M_j(k_2) T_{\mu\nu}(-q)\rangle& = A(q^2,k_1^2,k_2^2)(\eta_{\mu\nu}\,q^2 -q_{\mu}q_{\nu})\\
\noalign{\vskip.2cm}
& +B(q^2,k_1^2,k_2^2)(-\eta_{\mu\nu}\,q^2+2\,q_{\mu}q_{\nu})+C(q^2,k_1^2,k_2^2)(-\eta_{\mu\nu}\,r^2+2\,r_{\mu}r_{\nu})
\end{aligned}
\end{equation}
where $q^{\mu}\equiv k_1^{\mu}+ k_2^{\mu}$ and $r^{\mu}\equiv k_1^{\mu}- k_2^{\mu}$. The conservation and
trace Ward identities  completely determine $A,B,C$.
The diffeomorphism Ward Identity (conservation of the energy-momentum tensor) determines the $B$ and $ C$
amplitudes in terms of the
two-point function of the moduli
\begin{equation}
Q(k^2)\equiv \langle M_i(k)M_j(-k) \rangle	
\end{equation}
evaluated at $k_1^2$ and $k_2^2$ respectively, while the Weyl transformation (trace of energy momentum)
Ward identity determines the $A$ amplitude
in terms of the trace of the energy momentum tensor, i.e. the anomaly which we denote by $\cal{B}$:
\begin{equation}
A(q^2,k_1^2,k_2^2)= -{{\cal B}\over q^2}
\end{equation}
Therefore the $B ,C$ part of the decomposition obeys the diffeomorphism
Ward identity and it is traceless as seen from the explicit decomposition.
That part contains through the two point correlator of moduli the logarithmic divergence. It
follows that due  to the very special
kinematical features of  $d=2$, the three point function of two moduli and one energy momentum tensor splits into a
non-anomalous  part and a completely explicit anomalous part represented by the $A$ amplitude.
We remark the $1 \over q^2$ structure in the anomalous part which is surprising, since a priori one would expect
singularities combining the three kinematical invariants.
Once this lowest correlator is understood it is easy to write the result for any number of energy momentum tensors and moduli
by simply making the result covariant in space-time  and using covariance under source reparametrizations for the moduli.
The result for the anomalous part of the effective action
which has the  correct Zamolodchikov
trace anomaly is:
\begin{equation}
W_a={1\over 4\,\pi}\int d^2 x\sqrt{g}\,{\cal G}{1\over\square}R
\end{equation}
where we have defined
\begin{equation}
{\cal G}=G_{i\bar\jmath}\,\p^\mu J^i\,\p_\mu\bar J^{\bar\jmath}
\end{equation}
This can be combined in a single expression with the Polyakov trace anomaly since they have the same structure, 
i.e. $1/\square$.

The supersymmetrization is now  straightforward by replacing the scalar curvature in the Polyakov anomaly  and the
Zamolodchikov anomaly with their superspace generalizations in a single linear combination.
The relative normalization is fixed by the linear combination selected through factorization for the gauge components since the
superspace curvature contains the gauge field
$A$ while the superspace Zamolodchikov anomaly contains the K\"ahler $U(1)$ field $\cal{A}$.
Then the anomalous part of the effective action in superspace is
\begin{equation}\label{Wd2}
W=-{c\over 48\,\pi}\int d^2 x\int d^4\theta E\Big(\bar R-{6\over c}\nabla^2 K\Big){1\over\square}\Big(R-{6\over c}\bar\nabla^2 K\Big)
\end{equation}
whose super-Weyl variation is
\begin{equation}\label{Ad2}
{\cal A}=\int d^2 x\,d^2\theta\, {\cal E}\,\Sigma\Big(-{c\over24\pi}{R}+{1\over 4\pi}\bar\nabla^2 K\Big)~+~\hbox{c.c.}
\end{equation}
Its component expansion is \eqref{defA}.
That  \eqref{Ad2} follows from \eqref{Wd2} can be checked using
\begin{equation}
\begin{aligned}
\delta R&=-\Sigma\,R+2\,\bar\nabla^2\bar\Sigma\\
\delta\nabla^2&=-\bar\Sigma\,\nabla^2\,,\qquad \square =\bar\nabla^2\,\nabla^2
\end{aligned}
\end{equation}
Invariance under a joint K\"ahler shift $K\to K+f+\bar f$ and super-Weyl transformations with $\Sigma={6\over c}f$
is also manifest.
Here $\Sigma$ is a chiral superfield which parametrizes super-Weyl transformations;
its lowest component is $\Sigma|=\s+i\,\a$ and $R$ is the curvature chiral superfield, whose top component
contains the Ricci scalar (also denoted by $R$) and the $U(1)_V$ field strength $F$.
The sources $J^i$ are chiral superfields
and $K$ is a real function of the sources, the K\"ahler potential on the conformal manifold. For further details on the
geometry of ${\cal N}=(2,2)$ supergravity we refer to \cite{Grisaru}.
The anomalous effective action in super-conformal gauge was given in \cite{Gomis}.
The symmetry under a joint K\"ahler shift and a correlated Weyl transformation,  which acts on the
anomaly polynomial, is promoted here to a symmetry of the anomalous part
of the effective action.
We remark that the anomalous part as we defined it through factorization, contains a local Weyl invariant piece
$\sim\int d^2 x\int d^4\theta E K^2 $. To this we should add the fully Weyl invariant nonlocal contribution.
There is an   additional freedom we have since the Weyl invariance is anomalous: one is allowed to add  local
Weyl nonivariant functionals of the external sources respecting all the other symmetries:
\begin{equation}
\int d^2 x\,d^2\theta\, {\cal E}\,H(J){R}~+~\hbox{c.c.} +\int d^2 x\,d^4\theta\,E I(J,\bar J)
\end{equation}

It is instructive to have the anomalous  effective action also in components:
\begin{equation}
\begin{aligned}\label{d=2:components}
W&=\int d^2 x\sqrt{g}\Bigg({1\over 4\,\pi}{\cal G}{1\over\square}R
-{1\over 4\pi}F{1\over\square}\nabla^\mu{\cal A}_\mu-{c\over 96\pi}\Big( R{1\over\square}R+F{1\over\square}F\Big)
-{1\over 8\pi}K\,R\\
\noalign{\vskip.2cm}
&\qquad\qquad\qquad\qquad\qquad+{3\over 2\pi c}\Big({\cal G}\,K-{\cal G}{1\over\square}{\cal G}-{1\over4}K\,\square K
-\nabla^\mu{\cal A}_\mu{1\over\square}\nabla^\nu{\cal A}_\nu\Big)\Bigg)
\end{aligned}
\end{equation}
where we have defined $F=\epsilon^{\mu\nu}F_{\mu\nu}$. Note that it contains the gauge field and the K\"ahler connection
only in the combination $\mathbb{A}_\mu={\cal A}_\mu+{c\over6}\epsilon_{\mu}{}^\nu A_\nu$.

The last term in \eqref{d=2:components} represents correlators of moduli which are
only induced through the factorization assumption.
The local term in the OPE which through factorization produced  the  $\cal{A}$ dependent terms above was defined in
terms of $G_{i \bar\jmath}$, but after translating it to the covariant formalism
we ended up with an explicit dependence on $\cal{A}_\mu$.
Since the field strength ${\cal F}$ corresponding to $\cal{A}_\mu$, which is the pull-back of the K\"ahler form,
contains only $G_{i \bar\jmath}$, it is clear that in order to recover
the original information we should impose a gauge invariance of $\cal{A}$.
Such a gauge invariance is induced by a K\"ahler shift $K\to K+f(J)+\bar f(\bar J)$ i.e.
\begin{equation}
\delta{\cal A}_{\mu}={i\over 2}\p_\mu (f-\bar f)
\end{equation}
We are therefore led to study the behaviour of the anomalous part of the effective action under a K\"ahler shift.
Since the effective action is by construction invariant under a joint transformation by a K\"ahler shift and a Weyl transformation with
$\Sigma= {6 \over c}f$, the transformation under a K\"ahler shift can easily be calculated, 
simply replacing in the anomaly calculation $\Sigma$ by $f$:

\begin{equation}
\sim\int d^2 x\,d^2\theta\, {\cal E}\,f(J){R}
\end{equation}

The result of the K\"ahler shift  can be absorbed in a
change in the local Weyl noninvariant term by $H(J)\to H(J)+f(J) $

In summary in this class of theories through factorization the local terms in the OPE produce  contributions to the
effective action consistent with $(2,2)$ supersymmetry, but the Weyl anomalous part of the effective action is
not invariant under a K\"ahler shift, its variation being local. 

\section{Conclusions}

In \cite{Gomis} the behaviour of the anomaly polynomial under a K\"ahler shift was studied.
In this section, for the discussion of the implications of factorization, we find it convenient to discuss the behaviour
under a K\"ahler shift of the effective action itself. 
In this way we are able to isolate universal features of the terms generated by factorization which are 
not invariant under K\"ahler shift.  

We will consider terms in  the effective action which
depend on the moduli through a K\"ahler potential $K$.
When the ${\cal N}=2$ theory is the
result of a compactification  from a six dimensional theory on a Riemann surface
and $K$ has an ab initio geometric meaning \cite{TY,STY}, this is the case for the full effective action.
For a generic ${\cal N}=2$ theory  $K$ is defined by the moduli trace anomalies
and therefore we are really discussing only  the Weyl anomalous part of the effective action.
One expects a  ``K\"ahler shift invariance" for the transformation

\begin{equation}
K(J,\bar J) \rightarrow K(J,\bar J)+f(J)+\bar f(\bar J)
\end{equation}
This transformation induces on the pulled back universal $U(1)$ K\"ahler form a gauge transformation
\begin{equation}
{\cal A}_\mu\rightarrow {\cal A}_\mu+{i\over2}\p_{\mu}(f-\bar f)
\end{equation}
This K\"ahler shift can give a nonvanishing result. We will treat it as an anomaly with the understanding
that it originates just in those terms in the effective action which depend on $K$.
Then like for any other anomaly  one should look for  nontrivial solutions of the appropriate cohomological problem.
A partial solution  is given  by
\begin{equation}
\begin{aligned}\label{anomaly-superspace}
{\cal A}_f&={1\over 16\pi^2}\Big(\int d^4 x\,d^4\theta\,{\cal E}\,f\big(a'\,\Xi+b'\,W^{\a\b}W_{\a\b}\big)+\hbox{c.c.}\Big)
\end{aligned}
\end{equation}
or, in components:
\begin{equation}
\begin{aligned}\label{shift}
{\cal A}_f &
={1\over 32\pi^2}\int\!\! d^4 x\sqrt{g}\Bigg\lbrace -a'\,(f+\bar f)\left(E_4-{2\over3}\,\square R\right)
+(a'+b')\,(f+\bar f)\,C^{\mu\nu\rho\s}C_{\mu\nu\rho\s}\\
&\qquad-2\,(a'+b')\,(f+ \bar f)\, F^{\mu\nu}F_{\mu\nu}
+{1\over2}(a'+b')\,(f+\bar f)\,\tr\big(H^{\mu\nu}H_{\mu\nu}\big)
-2\,i\,b'\,(f-\bar f)\, F_{\mu\nu}\tilde F^{\mu\nu}\\
&\qquad+i\,b' \,(f- \bar f)\, R^{\mu\nu\rho\s}\tilde R_{\mu\nu\rho\s}
-{i\over 2}(a'-b')\,(f- \bar f)\,\tr\big(H_{\mu\nu}\tilde H^{\mu\nu}\big)
-4\,i\,a'\,\nabla^\mu A_\mu\,\square(f-\bar f)\\
&\qquad+8\,i\, a'\, A^\mu\Big(R_{\mu\nu}-{1\over3}R\, g_{\mu\nu}\Big)
\nabla^\nu(f-\bar f)-8\,a'\,F_{\mu\nu}\,A^\mu\,\nabla^\nu(f +\bar f)\Bigg\rbrace
\end{aligned}
\end{equation}
A third candidate
\begin{equation}
\int d^4 x\, d^8 \theta\, {E}(f+\bar f) K={1\over2} \delta_f \int d^4 x\, d^8 \theta\,E\, K^2
\end{equation}
is omitted, being cohomologically trivial in superspace.
It is an open question if there are possible additional  terms in the anomaly equation in which some
of the dependence on the ${\cal N}=2$ supergravity multiplet fields is replaced by a
dependence on some fields derived from $K$ itself. We will discuss this aspect in more detail below.

The K\"ahler shift anomaly  has the special feature that there are counterterms present with the same structure as
\eqref{anomaly-superspace} with arbitrary chiral coefficient functions and the anomaly shifts these coefficients.
These terms, though local, are chiral and therefore cannot cancel the K\"ahler shift produced starting from $K$ and therefore
this feature does not change the way we treat the anomaly.

The K\"ahler shift anomaly and Weyl anomalies have Wess-Zumino type consistency conditions which follow from the
commutativity of the two transformations:
\begin{equation}
\delta_f\delta_{\Sigma}W=\delta_{\Sigma}\delta_f W
\end{equation}
This equation fixes $a'$ in terms of the moduli contribution to the Weyl anomaly,
\begin{equation}
a'=-{1\over 24}
\end{equation}

This condition is equivalent to the invariance of the anomaly polynomial under a joint transformation with correlated 
$f$ and $\Sigma$ \cite{Gomis}.

On the other hand $b'$ is left unfixed since the expression it multiplies is Weyl invariant.
It follows immediately from the above discussion that if there are additional terms in the
K\"ahler shift anomaly polynomial they should be Weyl invariant since there is no term which
could match its Weyl variation.

We now come to the role of the local terms in the OPE.
Through factorization for every term in the effective
action involving $A_{\mu}$, we should get a corresponding term with ${1 \over 24 c}\cal{A}_{\mu}$ replacing $A_{\mu}$.
We will discuss  the implications for just the K\"ahler shift anomaly polynomial:

1) Replacing one $A_{\mu}$ by $\cal{A}_{\mu}$ in the Weyl anomalous generating functional generates under K\"ahler shifts 
terms with the same structure as those in \eqref{shift} which contain $f-\bar f$. 
For a general ${\cal N}=2$ theory in $d=4$ their normalization is 
however incompatible  with the relative normalization obtained by the Wess-Zumino condition.
This implies that  the local terms in the OPE, while contributing to the anomalous correlators, do not account for
the complete answer. For this we need to add the ordinary  contributions.

In Appendix B we describe the explicit check of a similar situation for the Maxwell supermultiplet:
the  Osborn anomaly is completely fixed by
${\cal N}=2$ supersymmetry in terms of the Riemann tensor computed from the Zamolodchikov metric. 
This was obtained as the sum of two terms,
one representing the factorized contributions of the local terms in the OPE and the other one the ordinary contribution.
Interestingly the two terms had even different index structures and
only their sum gave the Riemann tensor of the Zamolodchikov metric.

2) If we want to replace more than one $A_{\mu}$, we should limit ourself to the anomalous term involving three $A_{\mu}$ which
generate the $U(1)_{\cal R}$  chiral anomaly. Using factorization, terms depending on $A_{\mu}$ in the anomaly polynomial could
generate terms in the K\"ahler
shift anomaly polynomial. We will assume in the following discussion that the Weyl anomaly polynomial is ``complete"
i.e. the new anomalies suggested by factorization will appear only in the K\"ahler shift.
This can always be achieved by adding variations of local counterterms.
Then replacing two $A_{\mu}$
we get a new  term in the K\"ahler anomaly polynomial
$(f-\bar f){\cal F}_{\mu\nu}\tilde F^{\mu\nu}$. This term
should be made compatible with ${\cal N}=2$ supersymmetry,
i.e. obtained from an appropriate superspace expansion.
If the supersymmetrization turns out to be impossible,
the factorized contribution should be cancelled by an ordinary term.

Finally, by replacing all three $A_{\mu}$
we have the new term $(f-\bar f)\cal{F}_{\mu\nu}\tilde{\cal F}^{\mu\nu}$. This again should be supersymmetrized and the
previous discussion applies.
We remark that in order that the structures discussed under this point would appear, 
one needs at least four (real) moduli: otherwise the contributions vanish or are cohomologically trivial.
In summary in $d=4$, while the factorized local OPE contributions are needed, they seem to
act always together with the ordinary terms and their normalization therefore does not have unambiguous predictive power.

In $d=2$ the situation is different. Due to the specific two dimensional kinematical simplifications and $(2,2)$
supersymmetry, the Weyl anomalous part determines completely the K\"ahler shift anomalies in this component of the
effective action.
The  Weyl anomalous part of the effective action  can be separated unambiguously from the Weyl invariant part and it
depends on an explicit combination of the curvature superfield and the K\"ahler potential. The normalization of this combination
is determined by factorization and therefore the K\"ahler shift anomalies can be understood to follow entirely from Weyl
anomalies combined with factorization.
One cannot exclude of course that the Weyl invariant part of the effective action produces under the K\"ahler shift an
additional contribution with the same anomaly
structure but it is a consistent assumption that the Weyl invariant part is also K\"ahler shift invariant.

Finally we would like to comment on the possible role of local terms in the OPE in the conformal bootstrap. For theories with
extended supersymmetries the local terms should be included as additional couplings to the usual conformal blocks.
The constraints following from crossing symmetry should give interesting relations between the contributions of
local terms and the ordinary ones.

\bigskip

\noindent
{\bf Acknowledgements}\\
This work is supported in part by I-CORE program of the Planning and Budgeting Committee and the
Israel Science Foundation (grant number 1937/12) and by GIF -- the German-Israeli Foundation
for Scientific Research and Development (grant number 1265).
We thank the Galileo Galilei Institute for Theoretical Physics for hospitality; S.T. also acknowledges 
the hospitality extended to him at the Instituto de F\'{\i}sica Te\'orica (IFT)  in Madrid.  
We have benefitted from discussions with  D. Butter, Z. Komargodski, S. Kuzenko and M. Rocek.

\begin{appendix}

\section{Non-zero structure constant implies non-zero charge}

Consider in $d=4$ the correlator of a conserved current $J_{\mu}(z)$ with two dimension four operators
$M_{1}(x)$ and $M_{2}(y)$. We  assume that $M_1,M_2$ are not orthogonal to each other but we do not assume
anything about their charge under $J$. From conformal invariance the coordinate dependence is completely fixed
\cite{Petkos} for non-coinciding coordinates
\begin{equation}
\langle M_1(x)M_2(y)J_{\mu}(z)\rangle =c { 1 \over {(x-y)^6}}\left({1 \over {(z-x)^2}}{ (z-x)_{\mu}\over {(z-x)^4}}
- x\leftrightarrow y\right)
\end{equation}
except for the structure constant $c$.
The OPE between $J_\mu(z)$ and $M_{1}(x)$ can be extracted from the above correlator assuming that the representation holds
also when one coordinate approaches another. We put $x=0$ and $z$ infinitesimally
close to $0$ while $y$ is  kept fixed with  the component in the direction $\mu$ chosen to be $0$.
Then the OPE has the form
\begin{equation}
J_{\mu}(z) M_{1}(0)\sim c\,{z_{\mu} \over {z^4}}\,  M_{1}(0)
\end{equation}
Continuing to Minkowski space we obtain
\begin{equation}
T\big(Q(t) M_{1}(0)\big)\sim c\, \hbox{\rm sign}(t)\, M_{1}(0)
\end{equation}
where $Q$ is the charge operator $\int d^{3} z J_0(t,\vec{z})$ and $T$ is time ordering.
Considering the relation above for $t=\pm \epsilon$ we find
\begin{equation}
[Q, M_{1}(0)]\sim c\, M_{1}(0)
\end{equation}
i.e. $M$ is necessarily charged if the structure constant $c$ is not zero.

\section{The Maxwell case}

A simple toy model is four-dimensional ${\cal N}=2$ supersymmetric $U(1)$ gauge theory.
A useful reference is Appendix B of \cite{Baggio} whose notation we follow in this appendix.
The field content
are the gauge field, a $SU(2)_{\cal R}$ doublet of Weyl spinors and a complex scalar. There is also a
$SU(2)_{\cal R}$ triplet of auxiliary fields. They play no role in our analysis.
The action is
\begin{equation}
S=-{1\over g^2}\int d^4 x\left({1\over4}F_{\mu\nu}F^{\mu\nu}+{g^2\over 32\,\pi^2}\theta\, F_{\mu\nu}\,\tilde F^{\mu\nu}
+i\,\bar\lambda_i\,\bar\sigma^\mu\p_\mu\lambda^i
+\p_\mu\phi\,\p^\mu\bar\phi\right)
\end{equation}
The fermions carry $U(1)_{\cal R}$ charge $+1$ while the scalar has charge $+2$.
All other fields are neutral. The
$U(1)_{\cal R}$ current is therefore
\begin{equation}\label{jMaxwell}
j_\mu=-\bar\lambda_i\,\bar\sigma_\mu\lambda^i+2\,i(\phi\,\p_\mu\bar\phi-\bar\phi\,\p_\mu\phi)
\end{equation}
This theory has a complex modulus, i.e. an exactly marginal operator,
\begin{equation}\label{modulus}
M={i\,\pi\over 2}\left({1\over 8}F^+_{\mu\nu}F^{+\mu\nu}
+i\,\bar\lambda_i\,\bar\sigma^\mu\p_\mu\lambda^i-\bar\phi\,\square\phi\right)
\end{equation}
where $F^\pm=F\pm i\,\tilde F$. This operator is neutral under $U(1)_{\cal R}$ and one might
expect that the  $\langle M\,\overline M\,j_\mu\rangle$ correlation function vanishes. But this is not quite true
and it has, in fact, an imaginary part.
Note that the last two terms in \eqref{modulus} vanish on-shell. The reason why we are not allowed to
set these redundant operators to zero is supersymmetry. As we will show
they contribute in an essential way to the
three-point function. When inserted into a Feynman diagram they cancel a propagator, but the diagram still 
retains a nontrivial analytic structure. In the four-moduli correlator which we will compute below, the
redundant operators contribute in a similar way and their contribution is required in order to get the result 
which is consistent with supersymmetry. On the other hand it is also clear that their contribution to the 
two-point function is completely real and therefore for the Zamolodchikov metric only the gauge field 
part of the moduli is relevant. 

In this free field theory the $\langle M\,\overline M\,j_\mu\rangle$ correlation function is given by triangle
diagrams. Only the fermions and the scalar contribute and among the different possible contractions those where
the propagator between the $M$ and $\overline M$ insertions is cancelled, have an imaginary part.
This implies a local term
in the $M(x)\,\overline M(y)$ operator product expansion which is proportional to the current and the
$\langle j_\mu\,j_\nu\rangle$ two-point function is responsible for the logarithm.
\begin{center}
\vspace{-6cm}
\includegraphics[width=0.7\textwidth]{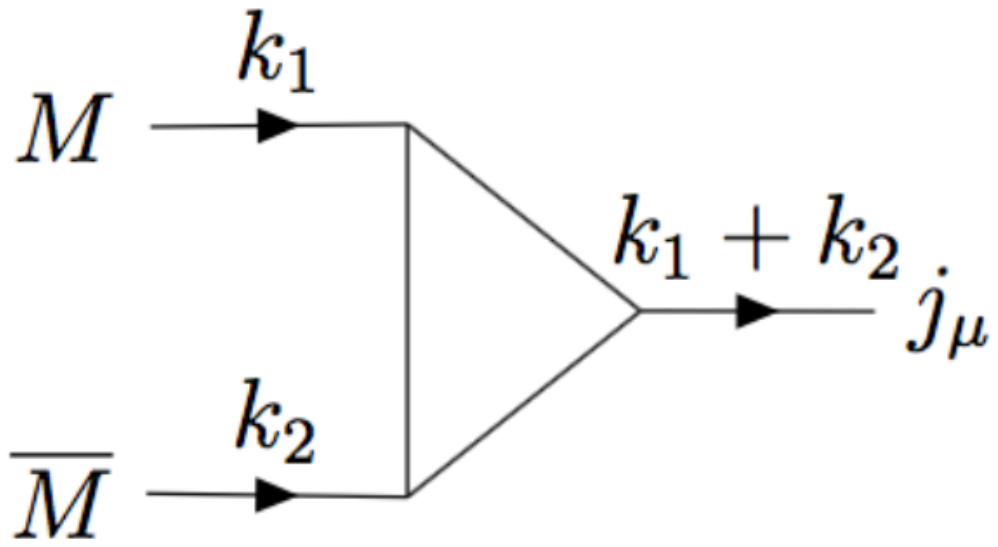}
\vspace{-6cm}
\end{center}
Explicit calculation of the one-loop triangle diagram gives
\begin{equation}
\langle M(k_1)\,\overline M(k_2)\,j_\mu(-k_1-k_2)\rangle
=-{1\over64}\big(q^2\,r_\mu-q\!\cdot\! r\,q_\mu\big)\log{\Lambda^2\over q^2}+\hbox{local}
\end{equation}
where $q=k_1+k_2$ and $r=k_1-k_2$.

Some comments/observations are in order:
(i) The correlation function is not gauge anomalous. This is consistent with \eqref{anpol4} where all moduli dependent
terms with $\a$ are cohomologically trivial.
(ii) It follows from the calculation that the logarithmic divergence is due to the cancelled propagators.
Those contractions where this does not happen, do not contribute.
(iii) The kinematical structure of the diagrams corresponds to $\int d^4 x\,F_{\mu\nu}\,{\cal F}^{\mu\nu}$.
This reflects the general structure of contact terms in this simplest example of a free theory.

We can also compute the logarithmically divergent part of the four-point function.
For the non-supersymmetric theory, where the modulus consists only of the spin one part in \eqref{modulus},
this was done by Osborn \cite{Osborn}.
His result cannot be cast into the form dictated by ${\cal N}=2$ supersymmetry, which contains
the target space Riemann tensor (cf. \cite{Gomis}, or the fourth line of eq. \eqref{anpol4}), which for the Zamolodchikov metric
$g_{\tau\bar\tau}={1\over 2\tau_2^2}$ is ${\cal R}_{\tau\bar\tau\tau\bar\tau}={1\over 4\tau_2^4}$.
Here $\tau$ is the single source in this case and $\tau_2$ its imaginary part. The difference between these two expressions
is proportional to
\begin{equation}\label{difference}
2\,(\nabla^\mu\tau\,\nabla_\mu\bar\tau)^2-5\, |\nabla^\mu\tau\,\nabla_\mu\tau|^2
\end{equation}
where now $\tau$ is the fluctuation around a constant value of the source and we have only kept terms
up to  ${\cal O}(\tau^4)$.
This difference must be accounted for by spin 0 and spin 1/2 contributions of $M$ and $\overline{M}$ via the cancelled
propagator argument. In each case, there are two Feynman diagrams which
contribute\footnote{The spin 0 and 1/2 parts of $M$ do not contribute
non-local parts to $M^3\,\overline M$ or $M^4$ correlators. As for the other orderings around the box digram, there are
always at least three cancelled propagators and therefore the Cutkosky rules give zero imaginary part and therefore
no logarithm.}
\begin{center}
\vspace{-9cm}
\includegraphics[width=1.0\textwidth]{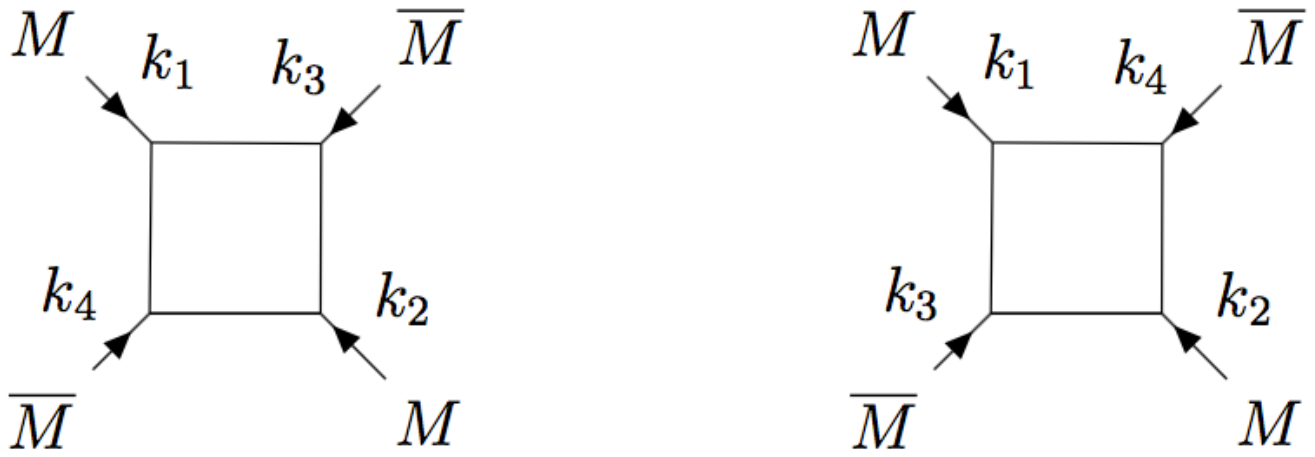}
\vspace{-10cm}
\end{center}
\noindent
For the spin zero part these two diagrams evaluate to
\begin{equation}
\begin{aligned}
&{\pi^2\over 384}\Big(m_1^2\,m_2^2+m_1^2\,m_3^2+m_1^2\,m_4^2+m_2^2\,m_3^2+m_2^2\,m_4^2+m_3^2\,m_4^2\\
&\qquad\qquad\qquad\qquad-(s+u)(m_1^2+m_2^2+m_3^2+m_4^2)+s^2+s\,u+u^2\Big)\log\Lambda^2+\hbox{finite}
\end{aligned}
\end{equation}
while for the fermions one computes
\begin{equation}
\begin{aligned}
&-{\pi^2\over 384}
\Big(2\, m_1^2\, m_2^2+2\, m_3^2\, m_4^2-m_1^2\, m_3^2-m_1^2\, m_4^2-m_2^2\, m_3^2-m_2^2\, m_4^2\\
&\qquad\qquad\qquad\qquad+(s+u)(m_1^2+m_2^2+m_3^2+m_4^2)-s^2-4 s u-u^2\Big)\log\Lambda^2+\hbox{finite}
\end{aligned}
\end{equation}
We have expressed the amplitudes in terms of an independent set of kinematical
invariants
\begin{equation}
m_i^2=k_i^2\,,\qquad s=(k_1+k_3)^2\,,\qquad u=(k_1+k_4)^2
\end{equation}
Their sum is proportional to
\begin{equation}
\begin{aligned}
&m_1^2\,m_2^2+m_3^2\,m_4^2-2(m_1^2\,m_3^2+m_1^2\,m_4^2+m_2^2\,m_3^2+m_2^2\,m_4^2)\\
\noalign{\vskip.2cm}
&\qquad\qquad+2(s+u)(m_1^2+m_2^2+m_3^2+m_4^2)-2(s^2+u^2)-5\,s\,u
\end{aligned}
\end{equation}
which is precisely the kinematical structure derived from \eqref{difference}. The overall normalization can
be fixed by an appropriate rescaling of the source $\tau$.

We remark that analysing the above diagrams in terms of the OPE we identify two additional local terms specific
to this model which contribute:
denoting by the $S(x)\equiv \phi(x)\bar\phi(x)$ the  dimension two scalar operator and by
$\kappa_\mu$ the (conserved) vector operator which differs from $j_\mu$ only by the relative sign between the
bosonic and fermionic contributions to \eqref{jMaxwell} and which can be shown to have vanishing two-point
function with $j_\mu$, we find
\begin{equation}
\begin{aligned}
M(x)\overline{M}(y)&\sim {\pi^2\over 32}\Big(\p_\mu^{(x)}\delta^4(x-y)\left({3}j^\mu(y)-\kappa^\mu(y)\right)
-\p_\mu^{(y)}\delta^4(x-y)\left({3}j^\mu(x)-\kappa^\mu(x)\right)\Big)\\
&\qquad\qquad+{\pi^2\over 8}\left(\square_{(x)} \delta^4 (x-y)S(y)-\p^\mu_{(x)}\delta(x-y)\,\p_\mu S(y)\right)
\end{aligned}
\end{equation}

\section{Free theory with moduli in $d=2$}

\medskip

We want to study the free (2,2) SCFT of a single twisted chiral  superfield $\Phi$
with unperturbed superspace action $\int\bar \Phi \Phi$. $\Phi=(\phi,\ps_+,\bar\psi_-,F)$ being twisted chiral
means that it satisfies
$\bar D_+ \Phi=D_- \Phi=0$ and the complex conjugate relations $D_+\bar\Phi=\bar D_-\bar\Phi=0$.
Here $D_\pm$ and $\bar D_\pm$ are flat superspace covariant derivatives.
We deform the theory by a chiral primary operator
\begin{equation}
M=\bar D_+\bar\Phi\,\bar D_-\Phi
\end{equation}
We then couple the deformed CFT to $U(1)_A$ supergravity.
If $J$ is the chiral source superfield, the deformed action is
\begin{equation}
\int d^2 x\, d^4\theta\,E\,\bar\Phi\Phi+\left(\int d^2 x\, d^2\theta\,{\cal E}\, J\, M+\hbox{c.c}\right)
=\int d^2 x\,d^4\theta\,E\,(1+J+\bar J)\Phi\bar\Phi
\end{equation}
A useful reference for flat $(2,2)$ superspace is Chapter 12 of \cite{Hori}. For curved superspace
we follow \cite{Grisaru}. With the help of results obtained there, we find the following component action
\begin{equation}
\begin{aligned}\label{compaction}
\int d^2 x\sqrt{g}\Big(&
-\phi\square\bar \phi+{i\over2}(1+J+\bar J)\bar\psi\,\gamma^\mu\mathrel{\mathop \nabla^{\leftrightarrow}}_\mu\psi
+(1+J+\bar J)\bar\psi\gamma^\nu\psi\big({\cal A}_\nu+\textstyle{1\over2} \epsilon_\nu{}^\mu A_\mu\big)\\
&\qquad\qquad+J\big(\p^\mu\phi\,\p_\mu\bar\phi-\epsilon^{\mu\nu}\p_\mu\phi\,\p_\nu\bar\phi\big)
+\bar J\big(\p^\mu\phi\,\p_\mu\bar\phi+\epsilon^{\mu\nu}\p_\mu\phi\,\p_\nu\bar\phi\big)\Big)
\end{aligned}
\end{equation}
Here we have set the gravitini to zero. The auxiliary scalar in the gravity multiplet drops out.
We have defined the Dirac spinor $\psi=(\psi_-,\psi_+)^T$ and
$J$ is now the lowest component of the source superfield.
The other components are set to zero as is the auxiliary field $F$ contained in $\Phi$; it vanishes on-shell.
$A_\mu$ is the $U(1)_A$ gauge field of the SUGRA multiplet
and ${\cal A}_\mu$ is the K\"ahler connection computed from the
potential\footnote{The relation with the more familiar K\"ahler potential for the metric on the upper half-plane,
$K=-\ln(\tau-\bar\tau)$
is established with the coordinate transformation $J=i\tau-{1\over2}$.}
\begin{equation}\label{KJ}
K=-\ln(1+J+\bar J)
\end{equation}

Note in \eqref{compaction} the relative factor ${1\over 2}$ in the coupling to the $U(1)$-current
$j^\mu\bar\psi\gamma^\mu\gamma\psi=\bar\psi\gamma^\nu\psi\epsilon_{\nu}{}^\mu$. It is ${c\over 6}$ for $c=3$,
the central charge
of the twisted chiral multiplet and we see that besides the coupling to gravity, the fermions
couple precisely to the combination of the $U(1)_A$ and the K\"ahler connection which was discussed in
Section \ref{d=2}.
From the action we can also read off the moduli operators as the coefficients of the sources. In a flat background
($g_{\mu\nu}=\eta_{\mu\nu},\,A_\mu=0$) they are
\begin{equation}
\begin{aligned}
M&=\p^\mu\phi\,\p_\mu\bar\phi-\epsilon^{\mu\nu}\p_\mu\phi\,\p_\nu\bar\phi+i\,\bar\psi\,\gamma^\mu\p_\mu\psi\\
\overline M&=\p^\mu\phi\,\p_\mu\bar\phi+\epsilon^{\mu\nu}\p_\mu\phi\,\p_\nu\bar\phi-i\,\p_\mu\bar\psi\,\gamma^\mu\psi
\end{aligned}
\end{equation}
The fermionic contribution vanishes on-shell i.e. it is redundant and will only contribute via the cancelled
propagator argument, already familiar from the discussion of the free Maxwell theory. The bosonic
part accounts for the ordinary contributions to correlators. As in the free Maxwell theory,
only this non-redundant part contributes to the logarithmic divergence of the
$\langle M\,\overline M\rangle$ two-point function and therefore to the Zamolodchikov metric.
If we expand the action around constant moduli, $J=\lambda+\delta J$ and compute
$\langle M\,\overline M\rangle$, we find that it is proportional to
$(1+\lambda+\bar\lambda)^{-2}$, from the normalization of the kinetic term of $\phi$ and the fact that the
one-loop diagram which computes it has two propagators.
This is the K\"ahler metric derived from \eqref{KJ}. Again as in the ${\cal N}=2$ Maxwell theory in $d=4$, the
redundant piece of $M$ is responsible for the non-vanishing of
$\langle M\,\overline M\,j_\mu\rangle$ and the cancelled propagator localizes the 
$M(x)\overline M(y)$ operator product on the $U(1)$ current.

If we integrate out $\phi$ and $\psi$, we recover the non-local effective action. It is easy to
integrate out the fermions. They can be rescaled to eliminate the $(1+J+\bar J)$ factor. What is left
are free fermions coupled to an external gauge field $\mathbb{A}_\mu\equiv{\cal A}_\mu+{c\over 6}\epsilon_{\mu}{}^\nu A_\nu$.
This leads to a term
\begin{equation}
\int d^2 x\, \p^\mu\mathbb{A}_\mu{1\over\square}\p^\nu\mathbb{A}_\nu
\end{equation}
in the effective action, in agreement with \eqref{d=2:components}.

It is special to this simple model that the microscopic action $\int d^4\theta e^{-K(J,\bar J)}\Phi\bar\Phi$
formally depends on the sources through the K\"ahler potential on the moduli space.
However the explicit expansion in components shows that due to the fact that the scalar fields without 
derivatives acting on them are not legal operators, 
the actual dependence on the sources does not necessarily 
reflect the coupling of the potential.
For this model therefore one can see explicitly that while the Weyl anomalous part of the action is defined by the K\"ahler potential 
with its potentially anomalous shift invariance, the couplings of the Weyl invariant part effectively do not contain anymore the 
K\"ahler potential.

\end{appendix}


\begin{thebibliography}{50}

\bibitem{Bzowski}
A.~Bzowski, P.~McFadden and K.~Skenderis,
``Implications of conformal invariance in momentum space,''
JHEP {\bf 1403} (2014) 111
[arXiv:1304.7760 [hep-th]],
``Scalar 3-point functions in CFT: renormalisation, beta functions and anomalies,''
JHEP {\bf 1603} (2016) 066
[arXiv:1510.08442 [hep-th]],
``Renormalised 3-point functions of stress tensors and conserved currents in CFT,''
  arXiv:1711.09105 [hep-th].

\bibitem{Prilepina}
A.~Dymarsky, K.~Farnsworth, Z.~Komargodski, M.~A.~Luty and V.~Prilepina,
``Scale Invariance, Conformality, and Generalized Free Fields,''
JHEP {\bf 1602} (2016) 099
doi:10.1007/JHEP02(2016)099
[arXiv:1402.6322 [hep-th]].

\bibitem{Nakayama}
Y.~Nakayama,
``On the realization of impossible anomalies,''
arXiv:1804.02940 [hep-th].

\bibitem{Kutasov}
D.~Kutasov,
``Geometry on the Space of Conformal Field Theories and Contact Terms,''
Phys.\ Lett.\ B {\bf 220} (1989) 153.

\bibitem{Gomis}
J.~Gomis, P.~S.~Hsin, Z.~Komargodski, A.~Schwimmer, N.~Seiberg and S.~Theisen,
``Anomalies, Conformal Manifolds, and Spheres,''
JHEP {\bf 1603} (2016) 022
[arXiv:1509.08511 [hep-th]].

\bibitem{DDI}
S.~Deser, M.~J.~Duff and C.~J.~Isham,
``Nonlocal Conformal Anomalies,''
Nucl.\ Phys.\ B {\bf 111} (1976) 45.

\bibitem{DS}
S.~Deser and A.~Schwimmer,
``Geometric classification of conformal anomalies in arbitrary dimensions,''
Phys.\ Lett.\ B {\bf 309} (1993) 279
[hep-th/9302047].

\bibitem{Friedan}
D.~Friedan and A.~Konechny,
``Curvature formula for the space of 2-d conformal field theories,''
JHEP {\bf 1209} (2012) 113
[arXiv:1206.1749 [hep-th]].

\bibitem{Wang}
J.~Gomis, Z.~Komargodski, H.~Ooguri, N.~Seiberg and Y.~Wang,
``Shortening Anomalies in Supersymmetric Theories,''
JHEP {\bf 1701} (2017) 067
[arXiv:1611.03101 [hep-th]].

\bibitem{Kuzenko}
S.~M.~Kuzenko,
``Super-Weyl anomalies in N=2 supergravity and (non)local effective actions,''
JHEP {\bf 1310} (2013) 151
[arXiv:1307.7586 [hep-th]].

\bibitem{deWit}
D.~Butter, B.~de Wit, S.~M.~Kuzenko and I.~Lodato,
``New higher-derivative invariants in N=2 supergravity and the Gauss-Bonnet term,''
JHEP {\bf 1312}, 062 (2013).
[arXiv:1307.6546 [hep-th]].

\bibitem{OsbornAnomaly}
H.~Osborn,
``Weyl consistency conditions and a local renormalization group equation for general renormalizable field theories,''
Nucl.\ Phys.\ B {\bf 363} (1991) 486.

\bibitem{Erdmenger}
J.~Erdmenger and H.~Osborn,
``Conserved currents and the energy momentum tensor in conformally invariant theories for general dimensions,''
Nucl.\ Phys.\ B {\bf 483} (1997) 431
[hep-th/9605009].

\bibitem{Grisaru}
S.~J.~Gates, Jr., M.~T.~Grisaru and M.~E.~Wehlau,
``A Study of general 2-D, N=2 matter coupled to supergravity in superspace,''
Nucl.\ Phys.\ B {\bf 460} (1996) 579
[hep-th/9509021].

\bibitem{STY}
N.~Seiberg, Y.~Tachikawa and K.~Yonekura,
``Anomalies of Duality Groups and Extended Conformal Manifolds,''
arXiv:1803.07366 [hep-th].

\bibitem{TY}
Y.~Tachikawa and K.~Yonekura,
``Anomalies involving the space of couplings and the Zamolodchikov metric,''
JHEP {\bf 1712} (2017) 140
[arXiv:1710.03934 [hep-th]].

\bibitem{Petkos}
H.~Osborn and A.~C.~Petkou,
``Implications of conformal invariance in field theories for general dimensions,''
Annals Phys.\  {\bf 231} (1994) 311
[hep-th/9307010].

\bibitem{Baggio}
M.~Baggio, V.~Niarchos and K.~Papadodimas,
``tt$^{*}$ Equations, Localization and Exact Chiral Rings in 4d $ \mathcal{N} $ =2 SCFTs,''
JHEP {\bf 1502} (2015) 122
[arXiv:1409.4212 [hep-th]].

\bibitem{Osborn}
H.~Osborn,
``Local couplings and Sl(2,R) invariance for gauge theories at one loop,''
Phys.\ Lett.\ B {\bf 561} (2003) 174
[hep-th/0302119].

\bibitem{Hori}
K.~Hori, S.~Katz, A.~Klemm, R.~Pandharipande, R.~Thomas, C.~Vafa, R.~Vakil and E.~Zaslow,
``Mirror symmetry,'' Clay mathematics monographs Vol 1 (2003)

\end{thebibliography}
\end{document}